\documentclass{ws-p8-50x6-00}

\def\PRC{{\em Phys. Rev.} C}

\begin{document}

%
%Equations with equation number followed by a lower case letter, e.g.. (6a)
%
\newcounter{saveeqn}
\newcommand{\alpheqn}{\setcounter{saveeqn}{\value{equation}}%
\setcounter{equation}{0}%
\renewcommand{\theequation}{\mbox{\arabic{saveeqn}\alph{equation}}}}
\newcommand{\reseteqn}{\setcounter{equation}{\value{saveeqn}}%
\renewcommand{\theequation}{\arabic{equation}}}
%%%
%%%%%%%%%%%%%%%%%%%%%%%%%%%%%%%%%%%%%%%%%%%%%%%%%%%%%%%%%%%%%%%%%

%%%%%%%%%%%%%%%%%%  BOLDFACE GREEK LETTERS   %%%%%%%%%%%%%%%%

\newcommand{\xbf}[1]{\mbox{\boldmath $ #1 $}}
%%%%%%%%%%%%%%%%%%
%%%%%%%%%%%%%%%%%%%%%%%%%%%%%%%%%%%%%%%%%%%
%

\title{Size and shape of baryons in \\
 a large N$_c$ quark model
}

\author{A. J. Buchmann}

\address{University of T\"ubingen, Institute for Theoretical Physics,
Auf der Morgenstelle 14, D-72076 T\"ubingen, Germany\\ 
E-mail: alfons.buchmann@uni-tuebingen.de}
\maketitle

\abstracts{Baryon charge radii and quadrupole moments  
are calculated in a quark model generalized to an 
arbitrary number ($N_c$) of colors. Several relations
among the charge radii and quadrupole moments are found.
In particular, the relation $Q_{p \to \Delta^+}=\frac{1}{\sqrt{2}}\, r_n^2$ 
between the neutron charge radius $r_n^2$ and the $p \to \Delta^+$ transition 
quadrupole moment $Q_{p\to \Delta^+}$ is shown to hold for physical 
baryons as well  in the large $N_c$ limit.}

\section{Introduction}

Charge radii ($r^2_B$) and quadrupole moments $(Q_B)$ 
are the lowest moments of the charge density $\rho$
in a low-momentum expansion. For example, for a member of the 
baryon decuplet with unit charge, one has up to $q^2$ contributions  
\be
\langle B \vert \rho(q) \vert B \rangle = 1 - \frac{q^2}{6} r_{B}^2 
- \frac{q^2}{6} Q_{B} + \cdots, 
\ee
where $q$ is the photon momentum transfer to the baryon. 
The first two terms arise from the spherically symmetric monopole 
part of $\rho$, while the third term is obtained from the quadrupole
part of the charge density. They characterize the total charge, 
spatial extension,  and shape of the system.

Baryon charge radii and quadrupole moments are closely related
consequences of the quark-gluon dynamics. They have been calculated in
different models of baryon structure, {\it e.g.}, an $N_c=3$
constituent quark model with two-body exchange currents~\cite{Buc97}.
Several new relations among the different $r_B^2$ and $Q_B$ have been
abstracted from these model calculations, some of which have later been
shown to be general consequences of the underlying symmetries and
dynamics of quantum chromodynamics~\cite{Mor99,Buc00}.

\section{Large N$_c$ Quark Model}

It is instructive to investigate the charge  radii and quadrupole
moments of baryons in a quark model in which the number of colors
$N_c$ is  not fixed to its physical value $N_c =3$ but allowed to take
on an  arbitrary odd integer value. Among the advantages of treating
$N_c$  as a parameter are:  \hfill \\
(i)  an expansion  in powers  of  $1/N_c$ of  the operator  structures
contributing  to a  given  observable  is obtained,  (ii)  there is  a
hierachy in importance of one-, two-, and three-quark operators.
% (iii) the $N_c=3$ constituent quark model and 
% the Skyrme model are seen to be different limits of the large $N_c$ 
% quark model.  

A multipole expansion of the baryon 
charge $\rho$ up to quadrupole terms 
leads to the following invariants  in spin-isospin space
\begin{eqnarray}
\label{eq:structure}
\rho & = & A \sum_i^{N_c} e_i -\frac{B}{N_c}\sum_{i < j}^{N_c}
 e_i \biggl [  2\xbf{\sigma}_i \cdot \xbf{\sigma}_j
-( 3 \xbf{\sigma}_{i\, z} \, \xbf{\sigma}_{j \, z} 
-\xbf{\sigma}_i \cdot \xbf{\sigma}_j ) \biggr ], 
\end{eqnarray}
where $\xbf{\sigma}_{i\, z}$ is the $z$-component of the Pauli spin
matrix of quark $i$, and $e_i$ is the quark charge.  The constants $A$
and $B$ in front of the one- and two-quark operators parametrize the
orbital and color matrix elements.  If the two-quark invariants are
generated by gluon exchange, there is a fixed ratio ($-2$) of the
factors multiplying the monopole (spin-scalar) and quadrupole
(spin-tensor) contributions to $\rho$.

Three-quark operators are neglected in the present work.

\section{Baryon Charge Radii }

According to Witten~\cite{Wit79}, baryons have charge radii
proportional to $N_c^0$ in leading order so that, {\it e.g.},
nonstrange baryons have the same size independent of their spin.
Here, we consider the $1/N_c$ corrections
associated with the spin-dependent two-body operators in 
Eq.~(\ref{eq:structure}).  The latter break SU(6) spin-flavor 
symmetry and lead to a splitting between octet and decuplet charge radii. 
This is analogous to the octet-decuplet mass splittings by spin-dependent
interactions~\cite{Jen95}.

The $N$ and $\Delta$ charge radii~\cite{Buc00} are the
matrix elements of the first two terms in Eq.~(\ref{eq:structure}) 
evaluated between large $N_c$ nucleon and $\Delta$
wave functions~\cite{KP84}:
\be
\label{crgeneric}
r_B^2 =
\frac{1}{Q} 
\left \{
A Q  - B 
\frac{4 J (J+1)  +  N_c (N_c +2) (2 Q -1 )}{N_c^2} + 6 B \frac{Q}{N_c},
\right \}, 
\ee
where $Q$ is the total charge and $J$ the total angular momentum of the baryon.

Specializing to $J=1/2$ for the nucleon states, 
one obtains (for neutral baryons there is no normalization factor $1/Q$) 
\vspace{0.3 cm}
\begin{eqnarray}
\label{crnucleon}
{r^2_p} & = &  {A} - {\displaystyle {B} 
\frac {(\,{N_c} - 1\,)\,(\,{N_c} - 3\,)}{{N_c}^{2}}} ,
\nonumber \\  
%%%%%%%%%%%%%%%%
%%%%%%%%%%%%%%%%%%%%%%%%%%%%%%%%%%%%%%%%%%%%%%%%%%%%%%%
{r^2_n}  &
= & \phantom{A}  \phantom{+} {\displaystyle 
{B} \, \frac {(\,{N_c} - 1\,)\,(\,{N_c} + 3\,)}{{N_c}^{2}}}. 
\end{eqnarray}
Similarly, from Eq.~(\ref{crgeneric}) we obtain for the $\Delta$ states
with $J=3/2$
\begin{eqnarray}
\label{crdelta}
r^2_{\Delta^{++}} 
& = & {A} - {\displaystyle \frac {3}{2}}\,B \, 
{\displaystyle \frac {(\,{N_c}^{2} - 2\,{N_c} + 5\,)}{{N_c}^{2}}},
 \nonumber \\  
%%%%%%%%%%%%%%%%%%%%%%%%%%%%%%%%%%%%%%%%%%%%%%%%%%%%%%%%%%%%%%%%%%
%%%%%%
r^2_{\Delta^+} & = & {A} - \,{\displaystyle 
{B} \, \frac {(\,{N_c}^{2} - 4\,{N_c} + 15\,)}{{N_c}^{2}}},
\nonumber \\  
%%%%%%%%%%%%%%%%%%%%%%%%%%%%%%%%%%%%%%%%%%%%%%%%%%%%%%%%%%%%%%%%%%
%%%%%%
r^2_{\Delta^0} & = & \phantom{A}  
\phantom{+} \,{\displaystyle {B} 
\, \frac {(\,{N_c} - 3\,)\,(\,{N_c} + 5\,)}{{N_c}^{2}}} ,
\nonumber \\
%%%%%%%%%%%%%%%%%%%%%%%%%%%%%%%%%%%%%%%%%%%%%%%%
%%%%%%%%%%%%%%%%%%%%%%%%
r^2_{\Delta^-} & = &  {A} - 3\, {\displaystyle {B}  
\frac{(\,{N_c}^{2} - 5\,)}{{N_c}^{2}}}. 
\end{eqnarray}

From these equations one readily derives that
\be
\label{rel1}
r_p^2 - r_{\Delta^+}^2 = r_n^2 - r_{\Delta^0}^2 
\ee
is valid for {\it all} $N_c$. Eq.~(\ref{rel1}) contains the SU(6)
symmetry-breaking effect due to the spin-isospin dependent ${\cal O}(1/N_c^1)$ 
correction in the baryon charge. Further relations and a
more complete treatment including the effect of three-body operators
can be found in Ref.~\cite{Buc00}.

\section{Baryon Quadrupole Moments}

Quadrupole moments measure the deviation of the baryon's 
charge density from spherical symmetry.
For the diagonal quadrupole moment of a nonstrange decuplet 
baryon with charge Q, total angular momentum $J$, and projection $J_z$ 
one finds
\be
\label{qmgeneric}
Q_B= {\displaystyle \frac{B}{2}} {\displaystyle \left \lbrack
\frac{[3\,{J_z}^{2} - {J}({J} + 1)]}{{J}({J} + 1)} 
\right \rbrack \, 
\frac{(4{J}({J} + 1) + {N_c}({N_c} + 2)(2{Q} - 1))}{{N_c}^{2}}}.
\ee
Eq.~(\ref{qmgeneric}) is the matrix element of the quadrupole part
of $\rho$ evaluated between large $N_c$ quark model
wave functions. Specializing to $J=J_z$ one gets for the 
different charge states of the $\Delta$,
\begin{eqnarray}
\label{qmdelta}
Q_{\Delta^{++}} & = & {\displaystyle \frac 
{2}{5}}\, B \, {\displaystyle 
\frac{[\,15 + 3\,{N_c}\,(\,{N_c} + 2\,)\,]}{{N_c}^{2}}} , \nonumber \\
%%%%%%%%%%%%%%%%%%%%%%%%%%%%%%%
%%%%%%%%%%%%%%%%%%%%%%%%%%%%%%%%%%%%%%%%%%%%%
Q_{\Delta^+} & = & 
{\displaystyle \frac {2}{5}}\,B \, 
{\displaystyle \frac {[\,15 + {N_c}\,(\,{N_c} + 2\,)\,]}{{N_c}^{2}}} ,
\nonumber \\
%%%%%%%%%%%%
%%%%%%%%%%%%%%%%%%%%%%%%%%%%%%%%%%%%%%%%%%%%%%%%%%%%%%%%%%%%%%%%
Q_{\Delta^0} & = & {\displaystyle \frac {2}{5}}\,B \,{\displaystyle 
\frac {[\,15 - {N_c}\,(\,{N_c} + 2\,)\,]}{{N_c}^{2}}} ,
\nonumber \\
%%%%%%%%%%%%%%%%%%%%%%%%%%%%%%%%%%%%%%%%%%%%%%%%%%%%%%%%%%%%%%%
%%%%%%%%%%%%%
Q_{\Delta^-} & = & {\displaystyle \frac {2}{5}}\,B \, 
{\displaystyle \frac {[\,15 - 3\,{N_c}\,(\,{N_c} + 2\,)\,]}{{N_c}^{2}}}. 
%%%%%%%%%%%%%%%%%%%%%%%%%%%%%%%%%%%%%%%%%%%%%%%%%%%%%%%%
%%%%%%%%%%%%%%%%%%%%
\end{eqnarray}

The $\Delta$ quadrupole moments satisfy the relations
\bea
\label{qmrel}
Q_{\Delta^{++}} + Q_{\Delta^-} & = &  Q_{\Delta^+} + Q_{\Delta^{0}}, 
  \nonumber \\
Q_{\Delta^{+}} + Q_{\Delta^-} & = &   2 \, Q_{\Delta^{0}}, 
\eea
which hold for {\it all} $N_c$. They are linear
combinations of the quadrupole moment relations obtained
with Morpurgo's general parametrization method~\cite{Buc01}. 
Only very weak assumptions, such as invariance of the strong interactions
under isospin rotations, are required to derive them.

The nondiagonal $N \to \Delta$ transition quadrupole
moment in the large N quark model is
\begin{equation}
\label{transqm}
Q_{N\to \Delta}= {\displaystyle 
\frac {1}{2}}\,B \, {\displaystyle 
\frac {\sqrt { 2\, (\,{N_c} - 1
\,)\,(\,{N_c} + 5\,)}}{{N_c}  }},
\end{equation}
both for the $p \to \Delta^+$ and the $n \to \Delta^0$ transition.
Additional results including an analysis of strange baryons and
three-body operators
will be published elsewhere~\cite{Leb02}.

\section{Relations Between Charge Radii 
and Quadrupole Moments }

Comparing Eq.~(\ref{transqm}) and Eq.~(\ref{crnucleon}), a relation between 
the neutron charge radius $r^2_n$ and the $p \to \Delta^+$ 
transition quadrupole moment $Q_{p \to \Delta^+}$ is obtained
\begin{equation}
\label{qmrel2}
Q_{p \to \Delta^+} = \frac{1}{\sqrt{2}} \, r^2_n,
\end{equation}
which is exact for $N_c=3$ and $N_c \to \infty$.
For intermediate values of $N_c$ 
the difference between left and right hand side
is ${\cal O}(1/N_c^2)$. Eq.~(\ref{qmrel2}) 
was originally derived in an $N_c=3$ constituent quark model
with two-body exchange currents~\cite{Buc97}. 
Recent experiments~\cite{Gra01,Got00} provide some evidence that it is 
satisfied within 20$\%$. 

The observables $r_n^2$ and $Q_{p \to \Delta^+}$ are closely related
because (i) in both cases one-quark operators do not contribute,
(ii) both observables are dominated by the two-quark ($B$) term,
(iii) the relative weight of the monopole (spin-scalar) and quadrupole 
(spin-tensor) parts in $\rho$ is uniquely determined
if these arise from gluon exchange.

The generalization of Eq.~(\ref{qmrel2}) to finite momentum transfers is
\begin{equation}
G^{p \to \Delta^+}_{C2}(q^2) = -\frac{3 \sqrt{2}}{q^2} G^n_{C0}(q^2). 
\end{equation}
Together with the result 
$G_{M1}^{p \to \Delta}(q^2)= -\sqrt{2} \, G_{M1}^n(q^2)$, it
can be used to predict the quadrupole (C2) over dipole (M1) ratio 
in $N\to \Delta$ electroexcitation from 
the elastic neutron form factor data~\cite{Gra01}. 
This agrees very well with C2/M1 data from $N\to \Delta$ 
electroexcitation~\cite{Got00}. 

Finally, the following relation between the
charge radii and quadrupole moments holds for {\it all} $N_c$:
\begin{equation}
\label{qmrel3}
Q_{\Delta^{++}}  + Q_{\Delta^+} + Q_{\Delta^0} + Q_{\Delta^{-}} =
r_p^2 - r_{\Delta^+}^2 + r_n^2 - r_{\Delta^0}^2. 
\end{equation}

\section{Summary}

The size and shape of baryons are different but related aspects
of the underlying quark-gluon dynamics. A large $N_c$ quark model
shows that there is a number of relations between 
baryon charge  radii and quadrupole moments 
that can be put to experimental tests.

\section*{Acknowledgments}
I thank the organizers for the invitation.
Some financial support by the DFG under title BU 813/3-1
 and the Institute for Nuclear Theory is gratefully acknowledged.
%Proceedings of Phenomenology of Large N QCD, World Scientific, 2002, 
%ed. R. F. Lebed, pg. 224.

\end{document}